\documentclass{aa}
\usepackage{graphicx}

\usepackage{longtable}

\begin{document}

\title{The spectroscopic orbits and the geometrical configuration of 
the symbiotic binary AR Pavonis\thanks{Based on observations taken at Complejo 
Astron\'omico El Leoncito (CASLEO), operated under an agreement between the Consejo 
Nacional de Investigaciones Cient\'{\i}ficas y T\'ecnicas de la Rep\'ublica Argentina, 
the Secretar\'{\i}a de Ciencia y Tecnolog\'{\i}a de la Naci\'on and the National 
Universities of La Plata, C\'ordoba and San Juan}}

\subtitle{}

\author{C. Quiroga\inst{1,2}, J. Miko\l ajewska\inst{3}, E. Brandi\inst{2,4}, O.
Ferrer\inst{1,2} and L. Garc\'{\i}a\inst{2}}

\offprints{C. Quiroga, \email{claudioq@muchi.fcaglp.unlp.edu.ar}}

\institute{ Consejo Nacional de Investigaciones Cient\'{\i}ficas y T\'ecnicas de la
Rep\'ublica Argentina (CONICET)
\and
   Facultad de Ciencias Astron\'omicas y Geof\'{\i}sicas - Universidad Nacional de La Plata
- La Plata - Argentina
\and
   Copernicus Astronomical Center- Warsaw - Poland
\and
   Comisi\'on de Investigaciones Cient\'{\i}ficas de la Provincia de Buenos de Aires
(CIC) - Argentina}

\date{Received  / Accepted }

\abstract{We analyze optical and near infrared spectra of intermediate and high resolution 
of the eclipsing symbiotic system AR Pavonis. We have obtained the radial velocity curves 
for the red and the hot component from the M-giant absorption lines and from the wings of 
H$\alpha$, H$\beta$ and He II $\lambda$4686 emission profiles, respectively. From the 
orbital elements we have derived the masses, $M_{g}=2.5M_{\odot}$ and 
$M_{h}=1.0M_{\odot}$, for the red giant and the hot component, respectively. 
We also present and discuss radial velocity patterns in the blue cF absorption spectrum as well as various emission lines. In particular, we confirm that the blue absorption lines are associated with the hot component. The radial velocity curve of the blue absorption system, however, does not track the hot companion's orbital motion in a straightforward way, and its departures from an expected circular orbit are particularly strong when the hot component is active. We suggest that the cF-type absorption system is formed in material streaming from the giant presumably in a region where the stream encounters an accretion disk or an extended envelope around the hot component. The broad emission wings originate from the inner accretion disk or the envelope around the hot star.
We also suggest that the central absorption in H profiles is formed in a neutral 
portion of the cool giant's wind which is strongly concentrated towards the orbital plane. 
The nebula in AR Pav seems to be bounded by significant amount  of neutral material in the 
orbital plane. The forbidden emission lines are probably formed in  low density ionized 
regions extended in polar directions and/or the wind-wind interaction zone.
}

\titlerunning{Spectroscopic orbits of the symbiotic binary AR Pavonis}
\authorrunning{C. Quiroga et al.}

\maketitle

\begin{table}
\caption[]{Log of the spectroscopic observations of AR Pav}
\label{table1}
\begin{tabular}[bottom]{lcccc}
\hline
\noalign{\smallskip}
Date	 
& 	
JD(2400000+)	 
& 	
Phase	 
& Detector &	
Range (\AA)\\
\noalign{\smallskip}
\hline
4/08/90 & 48117.7 & 0.966 & Z M & 5850-7200 \\
06/11/90 & 48201.5 & 0.105 & Z M & 5850 7100 \\
08/11/90 & 48204.5 & 0.110 & Z M & 4380-5050 \\
06/04/91  & 48352.8 & 0.355 & Z M & 5800-7100 \\
07/04/91 & 48353.8 & 0.356 & Z M & 4400-5050 \\
17/08/92 & 48852.7 & 0.182 & T CCD & 8300-9000 \\
19/06/95 & 49887.7 & 0.894 & R CD$^1$ & 4820-7800 \\
19/06/95 & 49887.8 & 0.894 & R CD$^1$ & 4820-7800 \\
12/08/95 & 49941.7 & 0.983 & R CD$^1$ & 4230-7300 \\
06/03/98 & 50879.9 & 0.535 & R CD$^1$ & 4320-7350 \\
07/03/98 & 50880.9 & 0.537 & R CD$^1$ & 4320-7350 \\
30/05/98 & 50963.9 & 0.674 & BC & 3900-5020 \\
07/09/98  & 51064.6 & 0.841 & R CD$^1$ & 5750-8750 \\
12/09/98 & 51069.6 & 0.849 & R CD$^2$ &4075-7100 \\
01/03/99 & 51238.9 & 0.129 & R CD$^1$ & 4625-7550 \\
26/05/99 & 51324.8 & 0.271 & R CD$^1$ & 4550-7550 \\
01/09/99 & 51422.6 & 0.433 & R CD$^1$ & 5700-8700 \\
02/09/99 & 51423.7 & 0.435 & R CD$^1$ & 4250-7300 \\
03/09/99 & 51424.5 & 0.436 & R CD$^1$ & 4250-7300 \\
25/03/00 & 51628.9 & 0.774 & R CD$^1$ & 4450-7550 \\
27/03/00 & 51630.9 & 0.778 & R CD$^1$ & 4450-7550 \\
29/07/00 & 51754.7 & 0.982 & R CD$^1$ & 4000-7100 \\
31/07/00 & 51756.7 & 0.986 & R CD$^1$ & 5500-8700 \\
\noalign{\smallskip}
\hline
\end{tabular}
\begin{list}{}{}
\item[] Z M: Reticon Z-Machine
\item[] T CCD: Thompson CCD (384x576 pixels)
\item[] R CD: REOSC echelle spectrograph in cross-dispersion
\item[] BC: Boller \& Chivens spectrograph
\item[1] images taken with bin factor=2
\item[2] image taken with bin factor=4

\end{list}\end{table}

\section{Introduction}

The symbiotic star AR Pavonis was discovered by Mayall (\cite{mayall37}) as an eclipsing 
binary with a period of 605 days. The eclipsed object is highly variable in both 
brightness and size (Andrews \cite{a74}). Thackeray \& Hutchings (\cite{th74}, hereafter 
TH74) based on extensive analysis of spectroscopic data spanning several orbital cycles 
proposed a binary model for AR Pav with the M3 III secondary filling its Roche lobe. They 
found that the eclipsed primary shows an emission O-type spectrum and  density and 
excitation in the eclipsed region increasing towards the central primary light source. 
TH74 derived masses of order $2.5\, \mathrm{M}_{\sun}$ and $1.2\, 
\mathrm{M}_{\sun}$, for the primary and secondary, respectively. Their estimate was based 
on analysis of the eclipse shape, and the primary mass function derived from radial 
velocities of almost exclusively  \ion{He}{i} emission lines. In particular,  they did 
not measure radial velocities for the cool giant. TH74 also found periodic changes in 
radial velocities of forbidden lines and cF-type  absorption lines, which they attributed 
to a stream and a region where the stream falls on a ring/disk surrounding the primary. 
Kenyon \& Webbink (\cite{kenyon84}) revised this model involving  an accretion disk 
around a main sequence star as the primary light source.

Bruch et al. (\cite{bruch94}) observed dramatic variations in the
visual light curve between eclipses which are not easily explained by the TH74 model.
They suggested variations in the mass transfer rate from the red giant to the hot 
component as the most satisfying explanation.
Skopal et al. (\cite{skopal00})
noted that the AR Pav light curve has a characteristic shape similar to those
of dwarf novae in quiescent phase; whereas during the active phases the profile
of the light curve changes considerably.

The first orbital solution for the cool giant has been recently obtained by Schild et al. 
(\cite{schild01}, hereafter S01). They also measured the rotation velocity of the giant, 
and, assuming co-rotation,  its radius. Based on this radius combined with the known 
spectral type of the giant and the corresponding effective temperature S01 estimated the 
giant's luminosity and distance to AR Pav, as well as from the position in the HR 
diagram the giant's mass. Their binary mass function then resulted in masses of $2\, 
\mathrm{M}_{\sun}$ and $0.75\, \mathrm{M}_{\sun}$ for the cool giant and its hot 
companion, respectively.

In this paper we present new radial velocity data for the cool giant and the emission
lines collected between 1990 and 2000. From this we derive new spectroscopic orbits for
both components, and discuss the velocity patterns in various emission lines.

\section{Observations}

Spectroscopic observations were performed with the 2.15 m telescope of Complejo
Astron\'omico el Leoncito, CASLEO (San Juan, Argentina). During 1990 and 1991 medium-low
resolution spectra were taken with a Boller \& Chivens Cassegrain spectrograph using a
photon-counting Reticon, called Z-Machine ($\lambda/\Delta\lambda$=4100 and 2700 in the
filters blue and  red respectively),  and a Thompson CCD of 384x576 pixels
($\lambda/\Delta\lambda$=2600) in 1992. Since 1995, high resolution spectra were obtained
with a REOSC echelle spectrograph using a Tek CCD 1024x1024 pixels
($\lambda/\Delta\lambda$=15000), except in May 1998 when a spectrograph Boller \& Chivens
was used. Table~\ref{table1} shows a log of the observations.

Barb\'a et al. (\cite{barba92}) describe details
of the Z-Machine, acquiring and reducing data procedures. The CCD data were
reduced with IRAF\footnote{IRAF is distributed by the National Optical
Astronomy Observatories, which is operated by the association of Universities
for Research in Astronomy, INC., under contract to the National Science
Foundation} packages, CCDRED and ECHELLE and all the spectra were measured
using the SPLOT task within IRAF.

To obtain the flux calibration, standard stars from Stone \& Baldwin (\cite{stone83}) and
Baldwin \& Stone (\cite{baldwin84}) were observed each night. A comparison of the spectra
of the standards suggests that the flux calibration errors are about 15 per cent for the
Z-Machine and 20 per cent in the central part of each order for the REOSC echelle images.

In the determination of the hot component orbit, we have included the
H$\alpha$ and He II$\lambda$4686 profiles measurements from a spectrum taken by Van
Winckel (\cite{winckel93}), in July 1988 and corresponding to phase $\phi=0.72$.

\section{Analysis and discussion}

\subsection{Emission line profiles}

\begin{figure*}
\centering
\includegraphics[width=17cm]{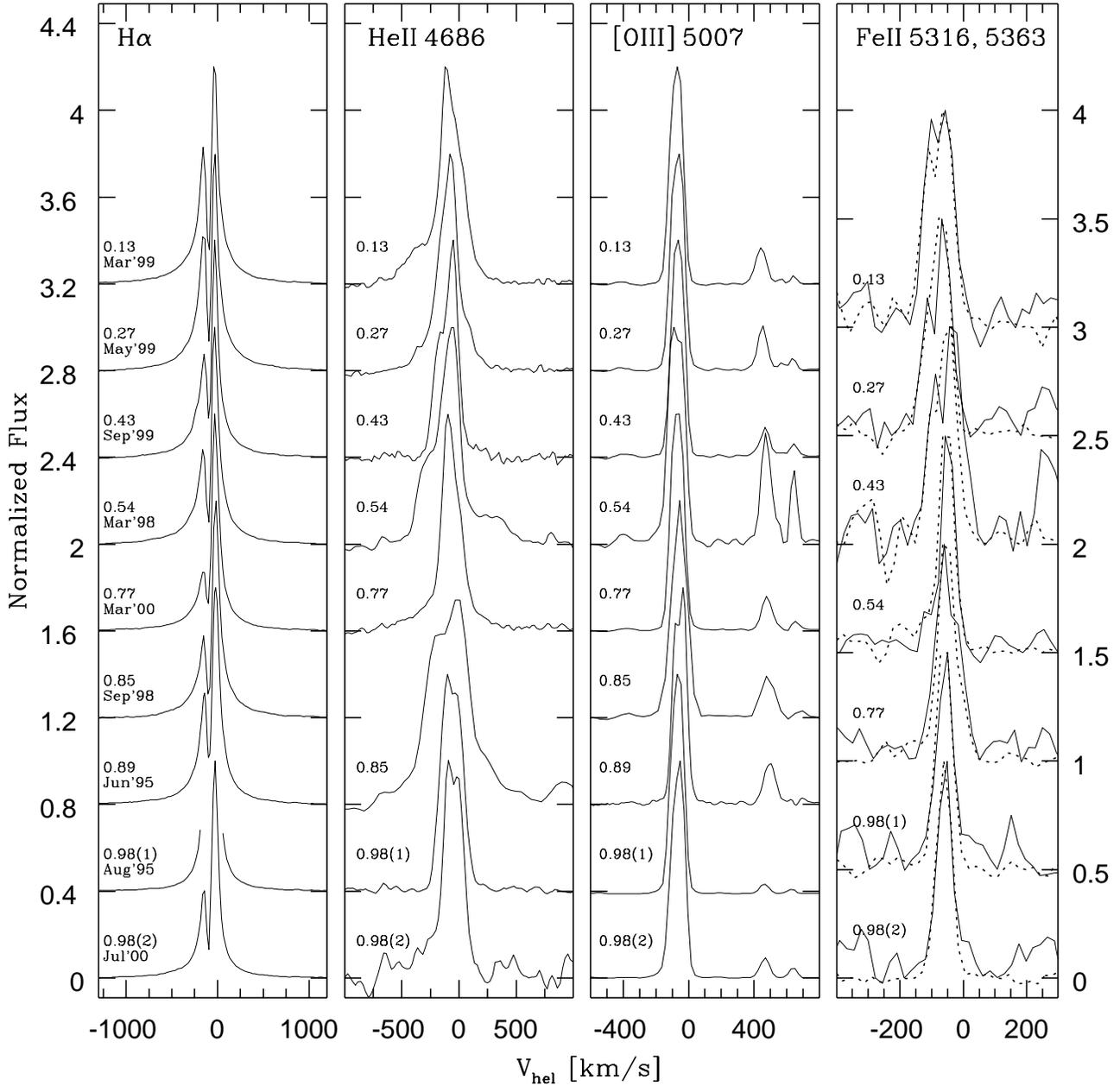}
\caption{Emission line profiles in AR Pav. The continuum was subtracted and each profile
was normalized to maximum intensity. The profiles are shifted vertically by 0.5 
(\ion{Fe}{ii}) and 0.4 (other lines), respectively, for
better display.  H$\alpha$ profile corresponding to August 95 is saturated. In the right panel, solid line shows Fe II $\lambda5316$ and dashed line shows Fe II 
$\lambda5363$. Only high resolution spectra are shown.}
\label{profiles}
 \end{figure*}

\begin{figure} 
\resizebox{\hsize}{!}{\includegraphics{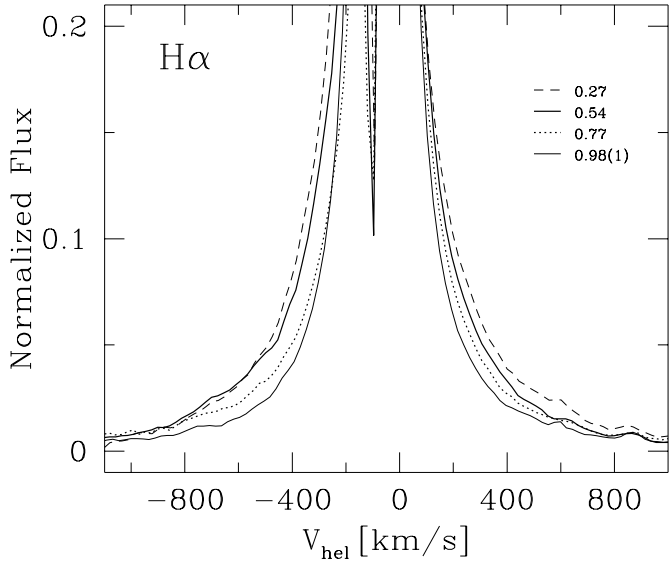}} 
\caption{
The wings of H$\alpha$ at selected orbital phases. }
\label{wings} 
\end{figure}

The left panel of Fig.~\ref{profiles} shows the H$\alpha$ emission line profiles of AR Pav 
ordered with their orbital phase, although they are collected from more than one cycle. 
H$\alpha$ is strongest in phase 0.54, when the hot component is in front. The profiles are 
double-peaked with broad, $\mathrm{FW} \ga 1600\,\mathrm{km\,s^{-1}}$, wings. The red peak 
is stronger than the blue one in all phases. This can not be explained by motions in an 
optically thin medium. In such situation the profile of the second quadrature would be
reversed relative to the first quadrature. The profiles are apparently affected by self-
absorption. The largest  blue to red peak intensity ratio is observed around phases 0.13 
and 0.27, respectively, whereas the smallest value of this ratio corresponds to phase 
0.77. The double-peaked profiles have an average separation of $148 \pm 3\, 
\mathrm{km\,s^{-1}}$. The average radial velocity of the central absorption is $-96 \pm 
1\, \mathrm{km\,s^{-1}}$. In eclipse, H$\alpha$ profiles are considerably narrower than 
out of eclipse  {\bf (Fig.~\ref{wings})}. This suggest that  the broad H$\alpha$ wings are 
formed near to the hot component.

The shape of H$\beta$  profile is similar to that of H$\alpha$, but the blue 
to red peak intensity ratio is smaller than the same ratio for H$\alpha$, except for 
phases 0.98 to 0.13, when the relative intensities are similar. The wings are also 
narrower during eclipse. The double-peaked profiles have a average separation of $117 \pm 
3\,\mathrm{km\,s^{-1}}$, and the average radial velocity of the central absorption is $ -
90 \pm 2\,\mathrm{km\,s^{-1}}$.

We have H$\gamma$ profiles only in phases near to both conjunctions. The intensity of the
blue peak relative to the red one is smaller than for H$\beta$, except during eclipse. 
It seems to behave in the same way as H$\alpha$ and H$\beta$. The average radial
velocity of the central absorption is $ -87 \pm 3\,\mathrm{km\,s^{-1}}$.
The absorption components of the Balmer lines seem to show a positive progression
from H$\alpha$ to H$\gamma$.
Similar positive progression in radial velocities of the central 
absorption component of Balmer emission lines is often observed in symbiotic stars.
Since the absorption lines have slower decrement than the emission lines
the positive progression 
may be due to the decreasing effect of the emission border.

The second panel of Fig.~\ref{profiles} shows the \ion{He}{ii} $\lambda\, 4686$ 
emission lines
profiles. The line profile is double-peaked during eclipse and in phase 0.85. During
eclipse the blue peak is slightly stronger than the red one. In phase 0.85 the red peak
is stronger than the blue one. In the rest of the orbital phases they present
asymmetric profiles. The asymmetry is reversed in the first quadrature relative to
second one. However, in phase 0.13 the shape of the asymmetry is almost identical to
phase 0.77. The largest asymmetry occurs in phase 0.54.

Fig.~\ref{profiles} also shows the [\ion{O}{iii}] $\lambda\,5007$ 
emission line region. The permitted lines of \ion{He}{i} $\lambda\,5016$ and \ion{Fe}{ii} 
$\lambda\,5018$ are narrow and more or less symmetric, with FW $\sim 100-150\, 
\mathrm{km\,s^{-1}}$. The forbidden [\ion{O}{iii}] lines display FW $\sim 200\, 
\mathrm{km\,s^{-1}}$ and may have double peak structure with peak-to-peak separation of 
$\Delta v \sim 40\, \mathrm{km\,s^{-1}}$, and the blue to red peak intensity ratio of 
$\sim 1$. This structure is visible in our profiles at phases 0.54 and 0.89 for 
which we estimate $\Delta v \sim 40$ and $50\, \mathrm{km\,s^{-1}}$, respectively 
(Fig.~\ref{profiles}), as well as in the profiles presented by van Winckel et al. 
(\cite{winckel93}) and S01. 

In the high resolution spectra, also some \ion{Fe}{ii} emission lines appear 
as double-peaked and/or asymmetric at most orbital phases (the right 
panel of Fig.~\ref{profiles}). In 
particular, the profiles are clearly double-peaked near the first quadrature, and 
asymmetric (with a very 
weak central absorption occasionally present) at the second quadrature. The profiles seem 
symmetric and the central reversal is  missing during the eclipse whereas at $\phi \sim 
0.54$ a very faint blue peak (hardly discernible from the noise) can be present. The 
double-peaked \ion{Fe}{ii} profiles have separations between 47 to 65  $\mathrm{km\, s^{-
1}}$. In the spectra of June 1995, the \ion{He}{i} emission lines $\lambda\lambda$ 4922, 
5876 and 6678 show also a double-peaked profile with an mean separation of $34\, 
\mathrm{km\, s^{- 1}}$, whereas the \ion{He}{i} $\lambda\,5016$  and the \ion{Fe}{ii} 
$\lambda\,5018$ profiles published by van Winckel et al. (\cite{winckel93}) show a 
reversal at $\sim -93$ and $\sim -86\, \mathrm{km\,s^{-1}}$, respectively. It is 
remarkable that in both conjunctions the intensity of \ion{Fe}{ii} $\lambda\, 4923$ and 
$\lambda\,5018$ emission lines are comparable with the intensities of \ion{He}{i} 
$\lambda\, 4921$  and  $\lambda\,5015$, respectively. In other phases the ratio Fe/He is 
very small.

\subsection{Radial velocities}

To obtain the radial velocities of the red giant and the hot component, we have measured
spectral features associated with each one.

For the cool component, we have used M-type absorption lines corresponding to
\ion{Fe}{i}, \ion{Ti}{i}, \ion{Ni}{i}, \ion{Si}{i}, \ion{O}{i}, \ion{Zr}{i}, \ion{Co}{i}, 
\ion{V}{i}, \ion{Mg}{i} and \ion{Gd}{ii}. 
At some phases  we have also identified and measured the blue cF-type absorption lines corresponding to \ion{Cr}{ii}, \ion{Fe}{ii}, \ion{Ti}{ii} and \ion{Y}{ii} which are 
believed to be linked to the hot companion (TH74; S01).
The individual radial velocities were obtained by a Gaussian fit of the line 
profile, and a mean value was calculated for each spectrum. The resulting mean 
heliocentric velocities together with their standard errors and the number of 
lines used are given in Table~\ref{table2}.

In addition, for the study of the hot component we have determined the radial
velocities from the wings of H$\alpha$, H$\beta$ and \ion{He}{II}\,$\lambda\,4686$ 
profiles. They
would reflect the motion of the hot component if they are formed in the inner region of
the accretion disk or near to the hot component. In this way we have used a method
outlined by Schneider \& Young (\cite{schneider80}) and Shafter (\cite{shafter83}), which
was very successful in studies of cataclysmic binaries. This method consists of
convolving the data ($S$) with two identical Gaussian bandpasses ($G$) whose centers have
a separation of $2b$. The wavelength, $\lambda$, of an emission line in a spectrum
$S(\Lambda)$ is given by solving the equation
\[\int_{-\infty}^{+\infty}S(\Lambda)G(\lambda-\Lambda)d\Lambda = 0\]
where
\[G(x)=\exp[-(x-b)^{2}/2\sigma^{2}]-\exp[-(x+b)^{2}/2\sigma^{2}].\]
The choice of the parameters $b$ and $\sigma$ depend on the
characteristic of the spectra being analyzed, namely the emission line width and the
signal-to-noise ratio of the data. We refer to Shafter (\cite{shafter83}) for a more
detailed discussion of the influence of the choice of $b$ and $\sigma$  on the obtained
orbital solution. In any case they found that varying these parameters has no significant
effect on the resulting value of the system velocity, $\gamma$, whereas the
semi-amplitude of the radial velocity, $K$, significantly depends on the choice of $b$.
The resulting velocities are given in Table~\ref{table2}.

In order to reduce the noise in the extreme line wings, we have added all spectra taken 
in the same observing run. Fig.~\ref{profiles} and Fig.~\ref{wings} show the H$\alpha$ 
emission line profiles and wings, respectively,
of AR Pav ordered with their orbital phase. We have experimented with a range of values 
for $b$, and $\sigma$ fixed at 7\,\AA, for which we have obtained the circular orbit 
solutions (see below). Fig.~\ref{ksig} illustrates the dependence of the semi-amplitude, 
$K$, 
and its standard error, $\sigma_\mathrm{K}$ on  $b$ in the case of H$\alpha$. It is 
noticeable that $\sigma_\mathrm{K}$ increases sharply for $b$ larger than  $\sim 
10$\,\AA. We can attribute these large standard errors to the velocity measurements 
dominated by the noise in the continuum rather than by the extreme high velocity wings of 
the line profile. We have therefore adopted $b= 10$\, \AA (460 $\mathrm{km\, s^{-1}}$) as 
the best value  in the case of  H$\alpha$.  A similar analysis led us to choose 
$b=7$\,\AA\ ($430\, \mathrm{km\,s^{-1}}$) and $\sigma = 5$\,\AA\ for H$\beta$,  
and $b=5$\,\AA\ ($320\, \mathrm{km\,s^{-1}}$) and $\sigma = 3$\,\AA\ 
for \ion{He}{ii} $\lambda\,4686$, respectively.

\begin{figure} \resizebox{\hsize}{!}{\includegraphics{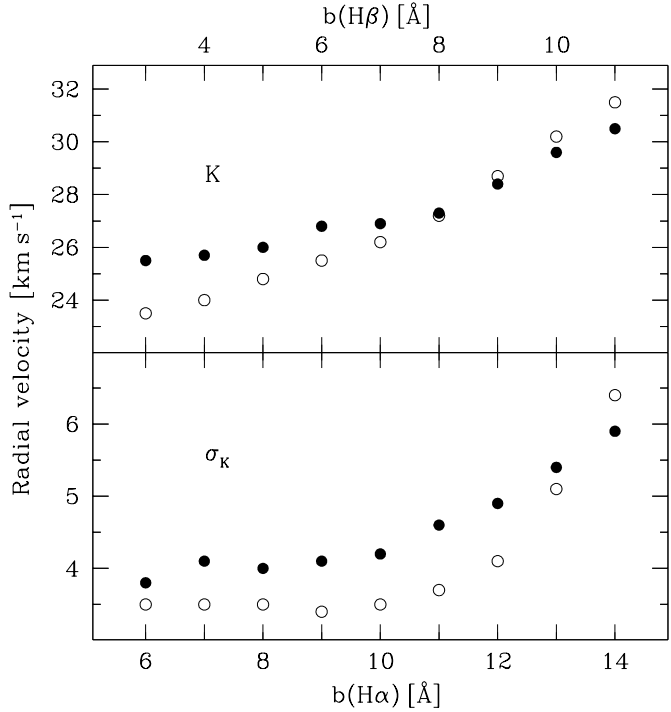}} \caption
{The orbital semi-amplitude $K$  and its standard error $\sigma_{K}$  for the wings of 
H$\alpha$ (dots) and H$\beta$ (open circles), respectively, 
as a function of the parameter $b$. The best estimation of $K$ is determined 
by the value of $b$ where $\sigma_{K}$ begins to sharply increase} \label{ksig} 
\end{figure}

We have also obtained the radial velocities of \ion{He}{i} $\lambda\lambda$ 4009, 4026, 
4121, 4144, 4388, 4471, 4713, 4922, 5016, 5048, 5876, 6678, 7065, 7281;  [\ion{O}{iii}] 
$\lambda\lambda\, 4363$, 4959, 5007;  [\ion{N}{ii}] $\lambda\, 6584$ and occasionally also 
$\lambda\, 5755$ emission lines as well as numerous \ion{Fe}{ii} emission lines by a 
Gaussian fit. Faint [\ion{Fe}{ii}] though present at some phases have not been measured. 
Table~\ref{table2} shows the average heliocentric radial velocities, their errors and 
number of lines used for each spectrum. The singlet and triplet series of \ion{He}{i} 
lines are 
shown separately.  In the case of double-peaked lines (see above) of Fe II the center 
velocity were used to derive 
the values in Table~\ref{table2}.

\begin{table*} \caption{Radial velocities of the giant absorption lines,  the blue cF-type 
absorption lines, permitted and forbidden emission lines, emission wings of \ion{H}{i} and 
\ion{He}{ii} lines$^{\star}$, and the central absorption component of H$\alpha$ (HA) in AR Pav} 
\label{table2} \footnotesize \begin{tabular}{lccccccccccc} \hline \noalign{\smallskip} 
JD/Phase & M--abs & cF-abs & \ion{Fe}{ii}  & \ion{He}{i}\,(s)  & \ion{He}{i}\,(t) & 
[\ion{O}{iii}] & [\ion{N}{ii}] & H$\beta$  & H$\alpha$ & \ion{He}{ii}$$  & HA \\ 
\noalign{\smallskip} \hline 
47356$^{\dagger}$/0.706 & & & & & & & & & -39 & & -96\\ 
47364$^{\dagger}$/0.719 & & & & & & & & & & -60 & \\ 
48118/0.966 & & & & -59\,(1) & -68$\pm$9\,(2) & & & & -75 & & \\ 
48202/0.105 & -59$\pm$3\,(5) & & & & & & & & -104 & & \\ 
48205/0.110 & & -79$\pm2$\,(9)& & -79$\pm$5\,(4) & -75$\pm$2\,(2) & -70$\pm$1\,(2) & & -81 
& & -99 & \\ 
48353/0.355 & & & -75\,(1) & -66$\pm$0\,(2) & -71$\pm$7\,(2) & -63$\pm$2\,(2) 
& & -87 & & -94 & \\ 
48354/0.356 & & & -80\,(1) & -69$\pm$5\,(2) & -72$\pm$2\,(2) & -65$\pm$2\,(2) & & & & & \\ 
48853/0.182 & -55$\pm$3\,(9)& & & & & & & & & \\ 
49888/0.894 & 
& -57$\pm2$\,(9) & -58$\pm$3\,(10)& -62$\pm$7\,(4) & -71$\pm$13\,(2) & -56$\pm$5\,(2) & : 
& -58 & -55 & & -93 \\ 
49942/0.983 & -76$\pm$2\,(13) & & -61$\pm$1\,(34) & -64$\pm$1\,(6) 
& -62$\pm$1\,(4) & -64$\pm$2\,(3) & -79$\pm$3\,(2) & -54 &-63& -58& s \\ 
50880/0.535 & -73$\pm$3\,(9) & -81$\pm1$\,(40)& -64$\pm$1\,(30) & -55$\pm$1\,(6) 
& -52$\pm$2\,(4) & -78$\pm$3\,(3) & -88$\pm$2\,(2) & -57 & -61 & & -105 \\ 
50881/0.537 & -72$\pm$2\,(9)&-81$\pm1$\,(43) & -65$\pm$1\,(32) & -55$\pm$1\,(6) 
& -50$\pm$2\,(4) & -77$\pm$2\,(3) & -85$\pm$1\,(2) & & & & -97 \\ 
50964/0.674 & & & & -60$\pm$9\,(5) & -61$\pm$6\,(3) & -87$\pm$4\,(3) & & -54 & & \\ 
51065/0.841 & -73$\pm$1\,(43) & & -47$\pm$2\,(10) & -50$\pm$1\,(2) & -48$\pm$1\,(2) & 
&-75$\pm$3\,(2) & & -36 & & -96 \\ 
51070/0.849 & &-46$\pm2$\,(21)& -41$\pm$2\,(24) & -47$\pm$1\,(5) & -44$\pm$2\,(5) 
& -59$\pm$1\,(3) & -76$\pm$2\,(2) &-26 & & -47 & -107 \\ 
51239/0.129 & -62$\pm$2\,(13) &  -81$\pm1$\,(22) & -79$\pm$1\,(22) &-81$\pm$2\,(5) 
& -78$\pm$4\,(3) & -71$\pm$1\,(2) & -91$\pm$4\,(2) & -79 & -81 & -98 & -101 \\ 
51325/0.271 & -56$\pm$2\,(18) & -85$\pm2$\,(14) & -79$\pm$1\,(24) &  -78$\pm$1\,(5) 
& -74$\pm$3\,(4) & -67$\pm$1\,(2) & -92$\pm$3\,(2) & -91 & -85 & -96 & -90\\ 
51423/0.433 
& -65$\pm$1\,(32)& & -73$\pm$1\,(10)& -72$\pm$2\,(2) & -61$\pm$3\,(2) & 
& -87\,(1) & & -70 & & -99 \\ 
51424/0.435 & & & -65$\pm$1\,(31) & -61$\pm$2\,(7) & -51$\pm$3\,(4) & -60$\pm$1\,(3) 
& -89\,(1) &-72 & & -76 & -86 \\ 
51425/0.436 & & &-70$\pm$1\,(22) & -66$\pm$2\,(7) & -61$\pm$4\,(4) & -71$\pm$1\,(3) 
& -89\,(1) & & & & -100 \\ 
51629/0.774 & -80$\pm$1\,(26) & -43$\pm1$\,(17) &-44$\pm$1\,(29) &  -47$\pm$1\,(5) 
& -44$\pm$2\,(4) & -68$\pm$0\,(2) & -79$\pm$3\,(2) & -34 & -41 & -33 & -93 \\ 
51631/0.778 & -79$\pm$3\,(21) & -42$\pm1$\,(15) & -47$\pm$2\,(27) & -48$\pm$1\,(5) 
& -44$\pm$3\,(4) & -69$\pm$1\,(2) & -85\,(1) & & & & -94 \\ 
51755/0.982 & -70$\pm$3\,(14)& & -60$\pm$1\,(33)& -63$\pm$2\,(6) & -60$\pm$2\,(5) 
& -63$\pm$1\,(3) & -82$\pm$4\,(2) & -56& -61 & -59 & -93 \\ 
51757/0.986 & -68$\pm$1\,(66) & & -62$\pm$2\,(13) & -59$\pm$1\,(2) 
& -61$\pm$2\,(2) & & -82$\pm$3\,(2) & & & & -94 \\

\noalign{\smallskip} \hline \end{tabular} \begin{list}{}{} \item [$\star$] Sum of the
spectra taken in the same observing run. \item [$\dagger$] Van Winckel(\cite{winckel93})
\item [s:] saturated emission \end{list} \end{table*}

\subsection{Spectroscopic orbits}

\begin{figure} \resizebox{\hsize}{!}{\includegraphics{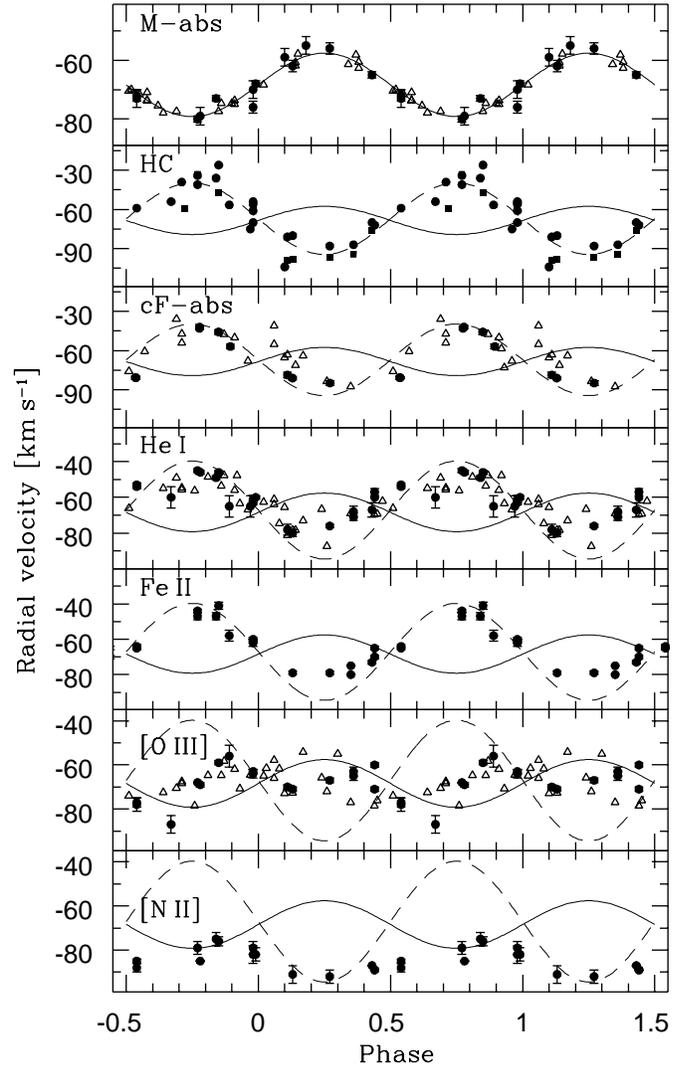}}
 \caption{Radial velocity 
data and circular orbital solution for AR Pav. Closed symbols represent our data, and 
triangles represent the data of S01 (M giant) and TH74 (cF absorption and emission lines). 
The solid line repeats the orbit of the M giant and the dotted line -- the hot component 
solution, respectively (a) M giant absorption. (b) Hot component (HC): filled circles 
correspond to the \ion{H}{i} wings, filled squares correspond to \ion{He}{ii} 
$\lambda$4686. (c)  cF absorption lines. (d) \ion{He}{i} emission lines. (e) \ion{Fe}{ii} 
emission lines. (f) [\ion{O}{iii}] emission lines. (g) [\ion{N}{ii}] emission lines.} 
\label{rvel} \end{figure} 

All radial velocities in Table~\ref{table2} except the central absorption of \ion{H}{i} 
vary with the 605-day period. Fig.~\ref{rvel} shows our radial velocity data corresponding 
to the M giant absorption lines, the \ion{H}{i} and \ion{He}{ii} emission wings,  and 
\ion{He}{i}, \ion{Fe}{ii}, [\ion{O}{iii}] and [\ion{N}{ii}] emission lines, respectively, 
phased with the photometric ephemeris given by Bruch et al. (\cite{bruch94}) 
\[\mathrm{JDMin} = 2\,420\,331.3+604.5\,E.\] Our radial velocity curve for the M giant is 
practically indistinguishable from that given by the data from S01 (and also plotted in 
Fig.~\ref{rvel}), and it  strongly suggests a circular orbit. It is also obvious from 
Fig.~\ref{rvel} that the cF absorption system, \ion{H}{i} and \ion{He}{ii} emission wings 
and the permitted (\ion{He}{i} and \ion{Fe}{ii})  emission lines are in antiphase with the 
M giant absorptions. 

In order to obtain the orbital parameters of AR Pav we have assumed a circular orbit as 
did S01, and adopted the very precise photometric ephemeris of Bruch et al. 
(\cite{bruch94}). The semi-amplitude of the respective radial velocity curve has been then 
fitted by a least squares method. In some cases we have also calculated elliptical orbital 
solutions with the orbital period forced to the photometric period of 604.5 d.
Table~\ref{table3} lists the resulting orbital solutions. The symbols have their usual 
meaning: $\gamma$ is the system velocity, $K$ is the orbital semi-amplitude, $e$ is the eccentricity, $\omega$ is the longitude of periastron, $T_0$ is the time of periastron 
passage, $f(M)$ is the mass function, $A \sin i$  is the fractional semi-major axis in 
astronomical units (AU), and $\Delta T$ is the time difference between orbital conjunction 
and photometric eclipse.  For the M giant, we have obtained both solutions using only  our 
radial velocities as well as for our radial velocities combined with the S01 data. 
Although an elliptical orbit fits the measured velocities slightly better than a circular one, the obtained eccentricity $e=0.12\pm0.04$ is only a $3\,\sigma$ result, and observations covering several more orbital periods would be necessary to confirm the reality of such an orbit. Following the arguments given by S01 we believe that a circular orbit provides a better description of the data, and such an orbit is also in agreement with the tidal theory.

\begin{table*} \caption{Orbital solutions for AR Pav} \label{table3} 
\begin{tabular}{lcccccccc} \hline \noalign{\smallskip} 
Component&$\gamma\, [\mathrm{km\,s^{-1}}]$ &$K\, [\mathrm{km\,s^{-1}}]$&$e$& $\omega$& 
$T_0^{(1)}$& $f(M)\, [M_{\sun}]$&$A \sin i$[AU]& $\Delta T^{(2)}$\\ \noalign{\smallskip} 
\hline 
M abs (our)&$-68.3\pm0.8$&$11.4\pm1.2$ & 0$^{(3)}$\\ 
M (our+S01)&$-68.4\pm0.3$&$10.8\pm0.5$& 0$^{(3)}$& & &$0.079\pm0.011$ &$0.60\pm0.03$& 0$^{(3)}$ \\
M (our+S01)$^{(4)}$&$-68.2\pm0.2$&$10.9\pm0.4$& $0.12\pm0.04$& $351\pm12$ & $50693\pm20$ & 
$0.079\pm0.009$&$0.60\pm0.02$& $12$ \\
cF (our+TH74)$^{(5)}$ & $-66.5\pm1.3$ & $27.6\pm2.9$ & $0.45\pm0.07$ & $273\pm11$ & 
$50899\pm21$& $0.94\pm0.40$ & $1.37\pm0.19$ & 34\\
cF (our+TH74)$^{(6)}$ & $-67.4\pm1.1$ & $22.8\pm1.7$ & $0.21\pm0.07$ & $307\pm20$ & 
$50936\pm31$& $0.69\pm0.18$ & $1.23\pm0.11$ & 4 \\
Wings (\ion{H}{i})&$ -65.2\pm1.9$ &26.3$\pm$2.7 &0$^{(3)}$&& &$1.14\pm0.32$ &$1.48\pm0.15$& 0$^{(3)}$ \\ 
Wings (\ion{H}{i})$^{(4)}$&$ -64.8\pm1.4$ &$25.8\pm1.8$ &$0.13\pm0.07$& $68\pm29$&$51726\pm45$ & $1.05\pm0.26$&$1.42\pm0.12$& 
$-7$ \\ 
Wings (\ion{He}{ii})&$-70.4\pm2.7$ & $28.9\pm3.9$ & 0$^{(3)}$& & &$1.51\pm0.70$ 
&$1.60\pm0.22$& 0$^{(3)}$ \\ 
Wings (\ion{H}{i}+\ion{He}{ii}) &$ -67.1\pm1.6$ & $27.4\pm2.3$ 
& 0$^{(3)}$& & &$1.29\pm0.35$ &$1.52\pm0.13$& 0$^{(3)}$ \\ 
\ion{He}{i} & $-63.3\pm1.8$ & $15.0\pm2.8$&0$^{(3)}$& & & && 0$^{(3)}$ \\ 
\ion{He}{i} (s)&$- 64.2\pm1.8$&$15.6\pm2.8$&0$^{(3)}$& & & & & 0$^{(3)}$ \\ 
\ion{He}{i} (t)&$-62.2\pm2.0$&$14.1\pm3.1$&0$^{(3)}$& & & & & 0$^{(3)}$ \\ 
\ion{Fe}{ii}&$-63.6\pm0.9$&$17.8\pm1.4$&0$^{(3)}$& & & &  & 0$^{(3)}$   \\ 
\noalign{\smallskip} \hline 
\end{tabular} \begin{list}{}{} \item [(1)] the time of periastron passage; \item [(2)] 
$\Delta T = T_\mathrm{conj}- T_\mathrm{eclipse}$; \item [(3)] assumed; \item [(4)] 
unconstrained solution; \item [(5)] unconstrained, weighted solution; \item 
[(6)] same as (5) for quiescent data only (see text).
\end{list} 
\end{table*}

Both the broad emission line wings, and the cF-type absorption lines show the highest (and practically the same) amplitude and the mean velocity almost identical with the red giant systemic velocity. The solutions for the broad emission wings of \ion{H}{i} and \ion{He}{ii}, respectively, agree within their 
respective errors which suggests that both they are formed in the same region near the hot component. 
An A- or F-type absorption system with radial velocity varying  in 
antiphase with the M giant absorption lines, similar to that identified in AR Pav, was seen during outbursts of two other 
symbiotic systems AX Per and BX Mon, and associated with the hot component  
(Miko{\l}ajewska \& Kenyon \cite{mk92a}, Dumm et al. \cite{dumm98}). This combined with 
the fact that the wings narrow during the hot component eclipse strongly suggests that 
both the broad emission wings and the blue (cF) absorption system trace the orbit of the hot component.
The only problem  is that any orbital solution for the cF absorption system leads to 
significant eccentricity (Table~\ref{table3}; TH74; S01) whereas the red giant orbit 
requires a circular (or nearly circular) orbit for the hot companion. 
Our orbital solution for the broad emission line wings, on the other hand, is consistent 
with a circular orbit.

\begin{figure} \resizebox{\hsize}{!}{\includegraphics{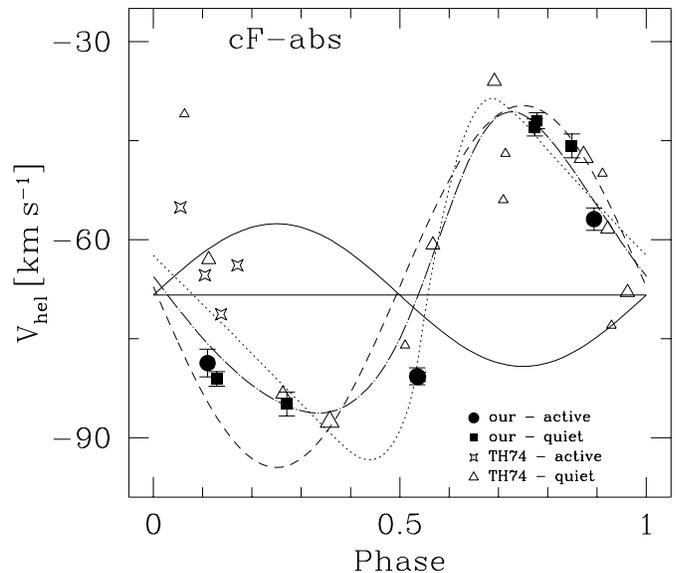}} \caption{The radial velocity curve for the cF shell component with our circular orbits of the red giant  (solid curve) and the broad emission wings (dashed curve), respectively. The horizontal line represents the system velocity, $\gamma=-68.4\, \mathrm km\,s^{-1}$. 
Open symbols (triangles and stars) represent the TH74 data, with sizes reflecting their 
weights. Filled symbols (squares and dots)  correspond to this work. We use different 
symbols for data obtained during activity and quiescence, respectively. The dotted curve 
gives an elliptical fit to all data points, the dot-dashed curve -- to quiescent data 
only. } \label{cfabs} 
\end{figure}

To analyze the cF absorption system we have combined our measurements with the 
TH74 data. Since our measurements are a factor of 2-4 more accurate then those of 
TH74, we have applied a weight 4 to our data, and weights 1-3 to the TH74 data 
(same as in their Table 1), respectively.  We have re-calculated the orbital 
solution for the combined data and found practically the same eccentricity, 
$e=0.45\pm0.07$, as $e\sim 0.45$ and 0.42 given by TH74 and S01, respectively, 
and $\gamma = -66.5\, \mathrm{km\,s^{-1}}$, closer to the red giant systemic 
velocity than earlier solutions. The spectroscopic conjunction occurs 34 days 
after the photometric eclipse.  S01 suggested that the cF absorption system 
only traces the hot component's motion whilst it is in front of the giant.
At other orbital phases, the cF absorption lines are affected by additional 
absorption due to this wind and/or a flow of material towards the hot 
companion, and the resulting radial velocities mimics an elliptical orbit. S01 
have also demonstrated that a circular orbit that is almost exactly in 
antiphase with the red giant results if only the radial velocity measurements 
between phase 0.25 and 0.75 ( the hot component in front) are 
employed. Unfortunately, our new measurements, in particular 2 points almost 
coincident at phase 
$\sim 0.5$ do not fit this restricted orbital solution of S01. Instead we have 
found that the cF absorptions may be affected by the hot component activity. This is illustrated in Fig.~\ref{cfabs}. 
In particular, all data points that force an elliptical orbit -- the TH74 points 
between phase $\sim 0$ and 0.2 as well as our two points near phase $\sim 0.5$ 
--  were obtained when AR Pav was very active (TH74; Skopal et al. 
\cite{skopal00b}).  It is interesting that an orbital solution omitting the 
TH74 data from the 1954 outburst as well as  our measurements obtained prior to 
JD\,$\sim 2\,451\,000$ (when according to Skopal et al. (\cite{skopal00b}) AR 
Pav  was very active) results in a much lower eccentricity, $e=0.21\pm0.07$ and 
the spectroscopic conjunction at the time of the photometric eclipse. 
Both these fits, however, match the data rather poorly, and the residuals for some points 
are larger then their measurement accuracy. This 
illustrates difficulties with interpretation of the radial velocity curve of 
the blue absorption system. 

We note here that similar complications with the blue absorption system occurred for a few other active symbiotic systems. In the case of symbiotic recurrent nova RS Oph, Dobrzycka \& Kenyon (\cite{dk94}) found the A-type absorption features $\sim 0.4\,P$ out of phase 
with the orbit of the  giant. Miko{\l}ajewska \& Kenyon (\cite{mk96}) failed to derive any 
radial velocity curve and orbital solution for the blue absorption system in Z And, 
whereas in CI Cyg the radial velocities of the F-type absorption system suggest formation 
region in material streaming from the giant near the hot component (Miko{\l}ajewska \& 
Miko{\l}ajewski \cite{mm88}).

Such problems are also common in the case of  W Ser stars (or hyperactive Algols), in 
which the accreting primary is embedded in an optically  thick shell with an A/F-type 
spectrum, and strong emission from circumstellar plasma is also present. 
We recall here that the radial velocity 
curves for the shell spectra in W Ser stars show often significant departure from that 
expected for a circular orbit required by the orbital solutions for their 
secondary components (Andersen et al. \cite{andersen88}).  It has been also found that such significant departures can be 
expected if the shell lines used to measure the radial velocity curve are formed in the 
outer $10-20\,\%$ of the disk (e.g. Andersen et al. \cite{andersen88}). In particular, the 
distorted radial velocity curve of the shell in SX Cas (Fig. 1 of Andersen et al. 
\cite{andersen88}) is very similar to the radial velocity curve of the blue absorption 
system in AR Pav. 

The possible correlation between AR Pav's activity and the departures of the cF 
absorption velocities from the circular orbit suggests that  this absorption 
system may be also affected by material streaming towards the hot component  presumably in 
a region where the stream encounters an accretion disk or an extended envelope around the 
hot component.
The outer disk and the stream impact region can be highly unstable and asymmetric. For 
example, the trailing 
side of the disk, where the gas  stream adds to the disk, can be brighter whereas the leading 
side is more extended. Such asymmetry may, at least  qualitatively, account for the 
deviations of the cF absorption velocity curve from the sinusoid expected for a circular 
orbit as well as for the behaviour of the optical light curves of AR Pav, which show 
relatively stable profile of eclipses during quiescence in contrast to very unstable 
profiles and changes in the position of the minima during activity (Andrews \cite{a74}; 
Bruch et al. \cite{bruch94}; Skopal et al. \cite{skopal00}).

Unfortunately, a detailed and 
proper treatment of the cF shell radial velocity curve is beyond the scope of our paper. 
This would require knowledge of the temperature and density profile in the shell/disk 
which is not 
available. The situation is further complicated by significant cycle-to-cycle variations 
of the cF spectrum.
In particular, spectroscopic observations a few orbital cycles and of 
good phase coverage in each cycle are necessary to distinguish between the cycle-to-cycle 
changes of the radial velocity curve and 
distortions due to other effects (e.g. contribution from stream, formation in outer disk, 
etc.).

The solutions for  the broad emission line wings are free from such complications and they 
are not affected by the system activity. We thus believe that the broad wings are formed 
very close to the hot star (in the inner disk or in a hot wind) and they directly trace 
the hot component's orbit.

Combing the semi-amplitudes of the M giant and the \ion{H}{i}+\ion{He}{ii} emission line 
wings for the circular orbit (Table~\ref{table3}) gives a mass ratio $q=2.5\pm0.3$, and 
component masses of $M_\mathrm{g}\sin^{3} i = 2.5 \pm 0.6\, \mathrm{M}_{\sun}$ and 
$M_\mathrm{h}\sin^{3} i= 1.0 \pm 0.2\, \mathrm{M}_{\sun}$, with the errors given by the 
errors of the $K_\mathrm{h}$ and $K_\mathrm{g}$ values, respectively. These values are 
somewhat larger than $M_\mathrm{g}\sin^{3} i = 1.9\, \mathrm{M}_{\sun}$ and 
$M_\mathrm{h}\sin^{3} i= 0.75\, \mathrm{M}_{\sun}$ derived by S01 mostly because 
our semi-amplitude for the M giant, $K_\mathrm{g}=10.8\,\mathrm{km\,s^{-1}}$,  is larger 
than the 
value $K_\mathrm{g}=9.6\,\mathrm{km\,s^{-1}}$ calculated by S01. We, however, note that 
our radial velocity curve  (Figure~\ref{rvel}) has better coverage at both quadratures.

The radial velocities of the central absorption component of \ion{H}{i} Balmer lines do 
not follow the orbital phase, and are blueshifted by $\sim 30\, \mathrm{km\,s^{-1}}$ 
relative to the system velocity. The absorption is presumably formed in the neutral 
portion of the cool component's wind. Moreover, it remains strong and practically 
constant over whole orbital cycle, which indicates that the nebula in AR Pav is probably 
bounded on all sides by significant amount of neutral material at least near the orbital 
plane. 
It is interesting that most symbiotic systems with double-peaked H$\alpha$ profiles 
have the main reversal on the blue side of the line center, strongly supporting the idea 
that this structure arises from self-absorption in the cool giant's wind (Ivison et al. 
\cite{ivison94}).

For a weak, unsaturated absorption the relationship between column density, 
$N_\mathrm{i}$, and the measured equivalent width, $EW$, is \[N_\mathrm{i} = \frac{1.13 
\times 10^{20}\,EW}{\lambda^2f_\mathrm{ij}}\mathrm{cm}^{-2},\] where $f_\mathrm{ij}$ is 
the oscillator strength of the transition $i \rightarrow j$, and wavelength $\lambda$ and 
$EW$ are given in \AA. From the measured $EW(\rm{H\alpha}) = 1.17 \pm 0.03\, \rm \AA$,  we 
estimate the lower limit for $N(\ion{H}{i}\,,n=2) \ga 4.4 \times 10^{12}$, and from 
$EW(\rm{H\beta}) = 0.66 \pm 0.06\, \rm \AA$ we obtain $N(\ion{H}{i}\,,n=2)\ga 3 \times 
10^{13}\, \mathrm{cm}^{-2}$, respectively. To estimate the total hydrogen column density, 
$N(\rm H)$,  the concentration of hydrogen atoms at the second excitation level has to be 
adopted which requires detailed knowledge on physical conditions in the gas. Assuming 
local thermodynamic equilibrium (LTE) and a  temperature, $\la 6000\, \mathrm{K}$, we 
obtain (from the Saha-Boltzman relation)  $N(\ion{H}{i}\,,n=2)/N(\rm H) \la 10^{-8}$, and 
the total hydrogen column density  $N(\rm H) \ga \mathrm{a\, few} \times 10^{21}\, 
\mathrm{cm}^{-2}$. Despite of all uncertainties involved in this estimate, the resulting 
hydrogen column density is consistent with the behavior of the shortwavelength  IUE 
spectra of AR Pav. These spectra show characteristic attenuation of the continuum flux for 
$\lambda \la 1500$\, \AA\  at any phase -- including SWP 5828 and SWP 13956 taken at 
orbital phases 0.27 and 0.37, respectively -- which can be accounted for by Rayleigh 
scattering in the neutral material  surrounding  the binary system. It has been shown that 
to produce a noticeable effect column densities in the range $N(\ion{H}{i}) \sim 10^{20}-
10^{24}\, \mathrm{cm}^{-2}$ are required (e.g. Schmid \cite{schmid97}). We thus conclude 
that both the strength of the central absorption in \ion{H}{i} Balmer lines and the 
behavior of the IUE SWP spectra are consistent with significant amounts of neutral 
material bounding the nebula in the orbital plane.

The radial velocity curves for \ion{He}{i} and \ion{Fe}{ii} emission lines are in 
antiphase with the cool giant, which suggests a possible association of these lines with 
the hot component. However, their semi-amplitude, 
$\sim 15$ and $\sim 18\,\mathrm{km\,s^{-1}}$ for \ion{He}{i} and \ion{Fe}{ii}, 
respectively, are a factor of $\sim 2$ lower than 
that  of the hot component, and their mean velocities are redshifted by 
$\sim 5\,\mathrm{km\,s^{-1}}$ with respect to 
the system velocity. Such discrepant amplitudes can be explained by a contribution from 
material between the two stars and in the neighborhood of the red giant which can both 
reduce the amplitude and shift the mean velocity (see also discussion in S01). We have not 
found any significant difference between the K values as well as $\gamma$ velocity derived 
for the singlet and triplet \ion{He}{i} emission series, respectively. There is also no 
difference between our radial velocity curve and the data from TH74, which means that the 
\ion{He}{i} formation region was the same at the two epoches.

The radial velocities of the forbidden lines show a more complicated pattern 
(Fig.~\ref{rvel}, bottom). The [\ion{O}{iii}] emission lines do not trace any of the two 
stellar components. Our  radial velocity data show a deep minimum around the phase $\sim 
0.7$ and another less marked minimum around the phase $\sim 0.2$, and the mean velocity, 
$-68\,\mathrm{km\,s^{-1}}$, very close to our red giant $\gamma$ velocity. A small 
amplitude periodic RV changes are also present in TH74 data. S01 suggested that the 
variability may be due to variable relative size of the blue and red peaks causing the 
line center of unresolved [\ion{O}{iii}] lines (such as [\ion{O}{iii}] $\lambda\,4363$ in 
TH74) 
to move bluewards and redwards. However, we have found the same RV values and changes by 
measuring the line centers and the wings, respectively, which in our opinion suggests that 
the changes are due to true motions.

Finally,  the [\ion{N}{ii}] emission lines show a small periodic changes roughly in phase 
with the hot component and the mean velocity blueshifted by $\sim 16\, \mathrm{km\,s^{-1}}$.

\subsection{The M giant, the hot companion and the geometry of AR Pav system}

AR Pav is well-established eclipsing system.  The eclipses are however not total as 
indicated by relatively blue colours during mid-eclipse, $B-V \sim 0.6$, much different 
from $B-V \sim 1.5$ expected for an M5 giant, and variable $UBV$ magnitudes during 
minimum which increase with the magnitudes outside eclipse (Andrews \cite{a74}; 
Hutchings et al. \cite{hutchings83}). Similarly,  the UV continuum flux measured from IUE 
spectra (cf. Fig. 5 of S01) shows a significant reduction near $\phi =0$, the continuum 
is, however, not fully eclipsed, and the shape of the light curve is asymmetric and 
complex. S01 estimated a limit $i \ga 79^{\circ}$ from the observed exit from 
eclipse for a point-like source and using the red giant radius derived from its 
rotational velocity.

An independent estimate of the orbital inclination can be made by comparing the apparent red giant radius
inferred from eclipses, $R_\mathrm{g,ecl}$, with its mean tidal radius, $R_\mathrm{t}$,  defined as the radius of a 
sphere with a volume equal to the Roche lobe volume. The
inclination is\[ \cos i \leq \sqrt{(R_\mathrm{t}/A)^2 - (R_\mathrm{g,ecl}/A)^2}.\]
Skopal et al. (\cite{skopal00}) obtained 
$R_\mathrm{g,ecl}/A = 0.30 \pm 0.02$\footnote{Note that this estimate is based only on the 
shape of light curve and it is independent of the model adopted in their paper.}, while
$R_\mathrm{t}/A = 0.46 \pm 0.01$ for $q = 2.5 \pm 0.3$ (Paczy{\'n}ski \cite{bep71}).
If the giant actually fills its tidal lobe, the orbital inclination is $i = 70^{\degr}
\pm 2^{\degr}$, so $M_\mathrm{g} \sim 3\, \mathrm M_{\sun}$ and $M_\mathrm{h} \sim 1.2\,
\mathrm M_{\sun}$. Otherwise these estimates give
a strong lower limit for $i$ and upper limits for the binary component masses.

Our orbital solution yields a binary separation of $A \sin i = 457 \pm 35\, \mathrm 
R_{\sun}$. For $i=90^{\circ}$, the giant radius derived from eclipses is then $0.3 A = 137 
\pm 20\, \mathrm R_{\sun}$, very close to the value $130 \pm 25\, \mathrm R_{\sun}$ 
derived by S01 from rotational velocity. The red giant radius extends to $\ga 65\, \%$ of 
the mean Roche lobe radius\footnote{The ratio of the observed red 
giant's radius to the  mean tidal radius, $R_\mathrm{t}$ is a better measure of the Roche 
lobe filling factor than the ratio of the giant's radius to the distance of the inner 
Lagrange point from the giant's center adopted by S01.} with the lower limit set by $i=90^{\circ}$. 
Based on similar estimate for the Roche lobe filling factor S01 concluded that the giant in AR Pav does not fill its tidal lobe. In our opinion, however, this result does not settle the problem.

Recently, Orosz \& Hauschildt (\cite{oh00}) have shown that rotational broadening kernels 
for Roche lobe filling (or nearly filling) giants can be significantly different from 
analytic kernels due to a combination of the nonspherical shape of the giant and the 
radical departure from a simple limb darkening law. As a result, geometrical information 
inferred from $v \sin i$ measurements of cool giants in binary systems, and in particular 
in symbiotic stars, are likely biased and must be treated with caution. In particular, 
among symbiotic systems with measured $v \sin i$ and accurate orbits, three of them, T 
CrB, CI Cyg and BF Cyg,  seem to contain tidally distorted giants as indicated by their 
red/near-IR light curves with evident ellipsoidal changes (Belczy\'{n}ski \& 
Miko{\l}ajewska \cite{bm98}; Miko{\l}ajewska \cite{mik01}; Miko{\l}ajewska et al. 
\cite{mik01g}). In CI Cyg, the rotational velocity of the giant is consistent with a 
synchronously rotating Roche lobe filling giant (Kenyon et al. \cite{km91}). In T CrB, 
however, the observed $v \sin i$ is by $\sim 20 - 30$\,\% lower than expected for such a 
giant (Belczy\'{n}ski \& Miko{\l}ajewska \cite{bm98}). A very low $v \sin i = 4.5\pm2\, 
\mathrm{km\,s^{-1}}$ has been also measured for the red giant in BF Cyg (Fekel et al. 
\cite{fekel01}). The problem is further complicated by the fact that the ellipsoidal 
variations can be hardly visible only in quiescent visual light curves, whereas  
systematic 
searches for such changes in the red and near-IR range where the cool giant dominates the 
continuum light are missing (Miko{\l}ajewska \cite{mik01}). In fact, a secondary minimum 
is also visible in the quiescent optical light curves of AR Pav (Fig.1 of Skopal et al. 
\cite{skopal00}). Observations in the red and near-IR range are, however, necessary to 
confirm whether this minimum is due to ellipsoidal changes of tidally distorted red giant.

Based on near IR TiO band depths, M{\"u}rset \& Schmid (\cite{murset99}) estimated a 
spectral type M5 III for the giant, whereas Allen (\cite{allen80}) deduced M6 III from the 
2.3-$\mathrm{\mu m}$  CO band. If the giant is similar to normal giants, then its 
effective temperature should be between  $T_{eff}(\mathrm{M5\,III})=3355$ K and 
$T_{eff}(\mathrm{M6\,III})=3240$ K (Richichi et al. \cite{rich99}), and luminosity, 
$L_\mathrm{g} \sim 1800 - 4700\, \mathrm L_{\sun}$, for $R_\mathrm{g} = 130\, \mathrm 
R_{\sun}$ and $R_\mathrm{g} = R_\mathrm{t} = 210\, \mathrm R_{\sun}$, respectively. 
Comparing the position of the red giant in the HR diagram with the evolutionary tracks for 
RGB/AGB stars, we find the red giant's mass range of $2\, \mathrm M_{\sun} \la  M \la 5\, 
\mathrm M_{\sun}$  (e.g. Vassiliadis \& Wood \cite{vassi93}; Schaller et al. 
\cite{schaller92}; Bessel et al. \cite{bessel89}), consistent with our dynamical mass 
estimate.

The radial velocity curve of the F-type absorption lines suggest that the shell 
spectrum is formed in the outermost accretion disk and/or the region where the stream 
encounters the inner accretion disk or an extended envelope around the hot component. The 
broad emission wings could be then formed in the inner accretion disk or the extended 
envelope around the hot source. TH74 came to similar conclusion based on the radial 
velocity behavior and the fact that the Balmer continuum served as a strong background for 
the cF absorption lines.
We can place some limit on the size of the inner disk radius assuming Keplerian rotation. 
Then we have 
\[R_\mathrm{d}/\mathrm{R_{\sun}} \approx 19 \times (M_\mathrm{h}/{\mathrm M_{\sun}}) 
(v_\mathrm{d}/100\, \mathrm{km\,s^{-1}})^{-2} \sin^2 i.\] 
In particular, the H$\alpha$ line wings follow the hot companion's orbit down to at least 
$\pm270\, \mathrm{km\,s^{-1}}$ ($b \ga 6\, \mathrm \AA$; Fig.~\ref{ksig}), which 
corresponds to  a Keplerian disk radius of $2.6$ and $2.8\, \mathrm R_{\sun}$, for $i=90^{\degr}$ 
and $70^{\degr}$, respectively.
These values are much lower than the radius of the eclipsed object, $R_\mathrm{h,ecl} \sim 
0.1\, A\, (46\, \mathrm R_{\sun})$, derived from optical eclipses (Skopal et al. 
\cite{skopal00}; Andrews \cite{a74}). Unfortunately, our method does not allow to measure 
accurately the wing positions for $b \la 6\, \mathrm \AA$ because of the presence of 
strong central absorption. We also note, that the projected Keplerian velocity at an outer 
disk rim with $R_\mathrm{d} \sim 46\, \mathrm R_{\sun}$ is  $\sim 128\, 
\mathrm{km\,s^{-1}}$, very   close to the average peak-to-peak separation in Balmer lines, 
$\sim 148$  for 
H$\alpha$ and $\sim 117\, \mathrm{km\,s^{-1}}$ for H$\beta$,  respectively. However 
neither the two peaks nor the central absorption trace the hot component orbital motion 
(see also Sec. 3.3). Thus we do not believe that the observed double-peaked structure 
originates in an outer accretion disk.   

The maximum velocity in the H$\alpha$ line profile, $\ga 800\, \mathrm{km\,s^{-1}}$, is 
consistent 
with a Keplerian velocity at a distance of $\la 0.3\,\mathrm R_{\sun}$, which sets an 
upper limit for the radius of the hot component. Although this value is an order of 
magnitude larger than the expected radius of a $1\,\mathrm M_{\sun}$ white dwarf, it falls 
in the range of the observed hot component radii in classical symbiotic systems (e.g. M{\"u}rset et al. \cite{mnsv91}). 
This result also rules out the presence of a main-sequence accretor in AR Pav.

The permanent presence of the strong central absorption in \ion{H}{i} Balmer emission 
lines, stationary with respect to the orbital motion, indicates that the nebula is bounded 
on all sides by significant amount of neutral material (see Sec. 3.3). On the other hand, 
the strong double-peaked [\ion{O}{iii}] lines require an extended low-density ionized 
region. 
A plausible solution to this apparent inconsistency is a bipolar ionization structure of 
the nebula, with a lower density towards the poles 
than in the orbital plane. We note here that similar strong [\ion{O}{iii}] lines together 
with strong central absorption in \ion{H}{i} emission lines were observed during late 
outburst phases and decline in CI Cyg and AX Per. Moreover, in AR Pav, the eclipses in 
\ion{H}{i} and \ion{He}{ii} lines are narrow with well-defined eclipse contacts (TH74; 
Quiroga et al., in preparation) likewise the eclipses in CI Cyg and AX Per during the 
outburst and its decline (Miko{\l}ajewska \& Kenyon \cite{mk92a,mk92b}) which indicates 
that they may arise in a non-spherical, perhaps bipolar, nebula or flow. Such a geometry 
is also consistent with the recent radio observations of CI Cyg (Miko{\l}ajewska \& Ivison 
\cite{mi01}). These bipolar structures could  be associated with the presence of an 
accretion disk in AR Pav and related symbiotic systems at least during their active 
phases. 

Even if the giant does not fill its Roche lobe, its wind is likely  focused 
towards the secondary and/or towards the orbital plane (Mastrodemos \& Morris 
\cite{mm98,mm99}; Gawryszczak, Miko{\l}ajewska \& R{\'o}{\.z}yczka 
\cite{gmr01}). In particular, 3-D hydrodynamic model calculations show that, 
in a system with orbital parameters similar to those 
of AR Pav, gravitational interaction of the cool 
giant's with the secondary all alone can produce an equatorial to polar density contrast as large as 
100-1000  giving rise to a bipolar geometry of the circumstellar nebula even in the absence of an accretion disk.

A bipolar geometry of the nebula of AR Pav can also account for the orbital 
phase-dependent changes of the [\ion{O}{iii}] and [\ion{N}{ii}] emission lines. 
In particular, such variations can be expected if these lines are formed in the hot 
component wind confined by the asymmetric dense wind from the red giant or in the 
wind-wind interaction zone (e.g. Walder \& Follini \cite{wf01}). 
The radial velocity curves for 
the [\ion{O}{iii}] and [\ion{N}{ii}] emission lines, respectively,  are not in phase, 
presumably because the two ions probe different regions, in particular  with different 
density (note that the forbidden line emission is the most effective if the electron 
density is near to the respective critical density).

\section{Concluding remarks}

We have obtained the circular orbits for both components of AR Pav from direct radial 
velocity measurements of the M-giant and the hot component. In particular, we have found 
that the hot component orbital motion can be successfully traced by radial velocities 
derived from  the broad emission wings of \ion{H}{i} Balmer lines and \ion{He}{ii} lines. 
Similar result has been recently obtained by Ikeda \& Tamura (\cite{ikeda00}; also Fekel 
et al. \cite{fekel01}) for V1329 Cyg, in which the \ion{H}{i} and \ion{He}{ii} broad 
wings appear to follow the hot component orbit.  With both radial velocity curves, we have 
obtained the orbital parameters and dynamical masses for each component. The masses 
obtained are $M_{g} \sin^3 i =2.5 \pm 0.6\, \mathrm M_{\odot}$ and $M_{h} \sin^3 i = 1.0 
\pm 0.2\, \mathrm M_{\odot}$ for the red giant and the hot component respectively.

Our M-giant orbit is in good agreement with that given by S01. We believe that the 
somewhat higher semi-amplitude derived from combined our and S01 velocity data  results 
from a better coverage of the final radial velocity curve near its minimum and maximum. 
Our improved mass function, $f(M_\mathrm{g}) = 0.079\,\mathrm M_{\sun}$, combined with the 
red giant's mass, $M_\mathrm{g}=2.0\, \mathrm M_{\sun}$, derived by S01 from evolutionary 
tracks yields $M_\mathrm{h}=0.87\, \mathrm M_{\sun}$, in agreement with our dynamical mass 
estimate. As in S01, our final component  masses are very different from earlier published 
values (TH74; Skopal et al. \cite{skopal00}). In particular, our upper limit for the hot 
component mass, $M_\mathrm{h} \la 1.2\, \mathrm M_{\sun}$,  is much lower than the $2.5\, 
\mathrm M_{\sun}$ and $4.5\, \mathrm M_{\sun}$ values found by TH74 and Skopal et al., 
respectively, and it does not require the presence of a main-sequence companion to the M 
giant.Moreover, our analysis (Sec. 3.4) practically rules out such possibility. Instead, we find that the hot component, like in most symbiotic stars, resembles central stars of planetary nebulae. However, we must stress that the hot component of AR Pav  seems to be one of the most 
massive among  thusfar studied symbiotic systems.

AR Pav is also one of the most active symbiotic systems. In particular, its 
optical light curve in addition to eclipses shows strong cycle-to-cycle 
variability (e.g. Andrews \cite{a74}; Bruch et al. \cite{bruch94}; Skopal et 
al. \cite{skopal00}) which is related to the presence and strength of the cF 
absorption spectrum and a blue continuum (TH74; Hutchings et al. 
\cite{hutchings83}). Such a blue A/F-type shell spectrum is typical for symbiotic 
systems during outbursts (active phases), in particular for those with multiple 
Z And-type eruptions, and it seems to be associated with the hot component 
(Miko{\l}ajewska \& Kenyon \cite{mk92a}, \cite{mk92b}). Usually, the blue shell
component disappears during a few years after the visual maximum, and a 
typical symbiotic system spends more time in quiescence than in activity 
(Miko{\l}ajewska \& Kenyon \cite{mk92a}). In the case of AR Pav however the 
F-type component remained visible, although weakened following the decline of 
the optical brightness,  throughout all the period 1953-1973 covered by the 
observations of TH74, as well as during our observations in 1990-2000. 
Moreover, a warm blue continuum was also present in 1980-1982 (e.g. Hutchings 
et al. \cite{hutchings83}); unfortunately there are not published optical 
spectra which could confirm the presence of any blue absorption system. Thus, 
the hot component of AR Pav, unlike the hot components in typical symbiotic 
systems, seems to remain active most of time.
There is also rather no doubt that this high activity is somehow related to  variable accretion processes.

Finally,  AR Pav shares many properties with W Ser stars (or hyperactive Algols). In 
particular,  both seem to have the accreting star embedded in an optically  thick shell 
with an A/F-type spectrum, and strong emission from circumstellar plasma. In the case of W 
Ser stars, there is strong evidence that  the shell spectrum is formed mainly in a 
geometrically and optically thick accretion disk seen nearly edge-on and in a gas stream 
(e.g. Andersen et al. \cite{andersen88} and references therein). Kenyon \& Webbink 
({\cite{kenyon84}) propose the same interpretation for the A/F- type shell spectrum 
observed in AR Pav, CI Cyg and a few other symbiotic systems during their active phases. 
Our analysis of radial velocity patterns in the F-type shell absorption as well as 
various emission lines in AR Pav supports such interpretation. The main difference 
between these symbiotic systems and W Ser stars is that the \ion{He}{ii} recombination 
lines, such as $\lambda\,1640$\,\AA\ and $\lambda\,4686$\,\AA\, are absent in W Ser stars 
whereas they are strong in AR Pav, CI Cyg and other related symbiotic stars (except during 
the maximum of the strongest optical outbursts). This seems to be related to different nature of the 
accretor hidden inside the disk, which in W Ser stars seems to be a main-sequence B-type 
star whereas in AR Pav and other symbiotics the accretor is less massive, $M_\mathrm{h} 
\la 1\, \mathrm M_{\sun}$, much hotter, and perhaps more compact. Both AR Pav and W Ser 
stars also have very complex and variable light curves, which can be accounted for by 
instabilities of a luminous disk  component, and/or variations in mass transfer and 
circumstellar gas in the system. A complex circumstellar envelope is also suggested by 
variable polarization in both AR Pav and W Ser stars (Brandi et al. \cite{brandi00} and 
Plavec \cite{plavec80}, respectively).

Although it is very premature to claim that all symbiotic systems with multiple outburst 
activity do have tidally distorted giants, and -- at least during active phase -- 
accretion disks, 
whereas the non-eruptive systems do not, the former seems be the case for CI Cyg, YY Her, 
and a few other active systems. The latter is also true in the case of two non-eruptive 
systems, V443 Her and RW Hya, which do not show any evidence 
for ellipsoidal variability in their near-IR light curves, and their near-IR magnitudes 
show only small fluctuations with the observational errors
(Miko{\l}ajewska et al. \cite{mik01g}).

AR Pav certainly deserves further studies. For example, near infrared monitoring can 
provide important information about the process of mass transfer and accretion by 
confirming or excluding a tidally distorted giant. The structure of the nebula and the 
hot continuum source may be reconstructed and tested during the eclipses. AR Pav has not 
yet been detected  in radio range, and the available upper limit $F_\mathrm{6\,cm} \la 14$ 
mJy (Belczy{\'n}ski et al. \cite{bm00}) is very far from the present detection  limits. 
Multifrequency radio observations would give us information about the mass loss and the 
geometry of ionized region.  We also desperately need good models of the structure and 
spectra of accretion disks and boundary layers for symbiotic stars.

\begin{acknowledgements} 

CQ is grateful to Nicolaus Copernicus Astronomical Center for hospitality and support 
during the partial preparation of this paper.
We thank Nidia Morrell for the acquisition of the 1999 May image. We wish to acknowledge 
the facilities and support offered by the staff of the CASLEO. 
We also thank the anonymous referee for valuable remarks and suggestions.
This research was partly 
supported by KBN research grants No. 2P03D 021 12 and No. 5P03D 019 20.\\

\end{acknowledgements}


\begin{thebibliography}{}

\bibitem[1980]{allen80}
Allen, A. D. 1980, MNRAS, 192, 531

\bibitem[1988]{andersen88}
Andersen, J., Nordstr{\"o}m, B., Mayor, M., Polidan, R.S. 1988, A\&A 207, 37

\bibitem[1974]{a74}
Andrews,P. J. 1974, MNRAS, 167, 635

\bibitem[1984]{baldwin84}
Baldwin, J. A., Stone, R. P. S. 1984, MNRAS, 206, 241

\bibitem[1992]{barba92}
Barb\'a, R., Brandi, E., Garc\'{\i}a, L., Ferrer, O. 1992, PASP, 104, 330

\bibitem[1998]{bm98} Belczy{\'n}ski, K.,  Miko{\l}ajewska, J. 1998, MNRAS, 296, 77

\bibitem[2000]{bm00} Belczy{\'n}ski, K.,  Miko{\l}ajewska, J., Munari, U., Friedjung, M. 
2000, A\&AS 146, 407

\bibitem[1989]{bessel89}
Bessell, M. S., Brett, J. M.; Wood, P. R., Scholz, M. 1989, A\&AS, 77, 1

\bibitem[2000]{brandi00} 
Brandi, E., Garcia, L.G., Piirola, V., Scaltriti, F., Quiroga, C. 2000, A\&AS, 145, 197

\bibitem[1994]{bruch94}
Bruch, A., Niehues, M., Jones, A.F. 1994 A\&A, 287, 829

\bibitem[1994]{dk94}
Dobrzycka, D., Kenyon, S.J. 1994, AJ 108, 2259

\bibitem[1998]{dumm98}
Dumm, T., M\"{u}rset, U., Nussbaumer H., et al. 1998, A\&A, 336, 637.

\bibitem[2001]{fekel01}
Fekel, F.C., Hinkle, K.H., Joyce, R.R., Skrutskie, M.F. 2001, AJ, 121, 2219

\bibitem[2002]{gmr01}
Gawryszczak, A.J., Miko{\l}ajewska, J., R{\'o}{\.z}yczka, M. 2002, A\&A, in press

\bibitem[1983]{hutchings83}
Hutchings, J.B., Cowley, A.P., Ake, T.B., Imhoff, C.L. 1983, ApJ, 275, 271

\bibitem[2000]{ikeda00}
Ikeda, Y., Tamura, S. 2000, PASJ, 52, 589

\bibitem[1994]{ivison94}
Ivison, R.J., Bode, M.F., Meaburn, J. 1994, A\&AS, 103, 201

\bibitem[1984]{kenyon84}
Kenyon, S.J., Webbink, R.F. 1984, ApJ, 279, 252

\bibitem[1991]{km91}
 Kenyon, S.J., Oliversen, N.A., Miko{\l}ajewska, J., et al. 1991, AJ, 101, 637

\bibitem[1998]{mm98}
Mastrodemos, N., Morris, M. 1998, ApJ, 497, 303

\bibitem[1999]{mm99}
Mastrodemos, N., Morris, M. 1999, ApJ, 523, 357

\bibitem[1937]{mayall37}
Mayall, M. W. 1937, Ann. Harv. Coll. Obs., 105, 49 1

\bibitem[2001]{mik01}
Miko{\l}ajewska, J. 2001, in Small-Telescope Astronomy on Global Scales, ed. B. 
Paczy{\'n}ski, W.P. Chen, \& C. Lemme, ASP Conf. Ser., 246, 167/astro-ph/0103496

\bibitem[2001]{mi01} Miko\l ajewska, J., Ivison, R.J. 2001, MNRAS, 324, 1023 

\bibitem[1992a]{mk92a}
Miko\l ajewska, J., Kenyon, S. J. 1992a, AJ, 103, 579

\bibitem[1992b]{mk92b} Miko\l ajewska, J., Kenyon, S. J. 1992b, MNRAS, 256, 177

\bibitem[1996]{mk96}
Miko\l ajewska, J., Kenyon, S. J. 1996, AJ, 112, 1659

\bibitem[1988]{mm88}
Miko{\l}ajewska, J., Miko{\l}ajewski, M. 1988,  in The Symbiotic Phenomenon, ed. J. Miko{\l}ajewska et al. (Kluwer, Dordrecht) 187 

\bibitem[2002]{mik01g}
Miko{\l}ajewska, J.,  Kolotilov, E.A., Shenavrin, V.I.,  Yudin, B.F. 2002, 
in B.T. Gansicke, K. Beuermann, K. Reinsch, eds, The Physics of Cataclysmic Variables and 
Related Objects, ASP Conf. Ser, in press 

\bibitem[1999]{murset99} 
M\"{u}rset, U., Schmid, H. M. 1999, A\&AS, 137, 473

\bibitem[1991]{mnsv91} 
M\"{u}rset, U., Nussbaumer, H.,  Schmid, H. M., Vogel, M. 1991, A\&A, 248, 458

\bibitem[2000]{oh00} Orosz, J.A., Hauschildt, P.H. 2000, A\&A, 364, 265

\bibitem[1971]{bep71}
Paczy{\'n}ski, B. 1971, ARA\&A, 9, 183

\bibitem[1980]{plavec80}
Plavec, M. J. 1980, in IAU Symp. 88, Close Binary Stars: Observations and Interpretation, 
eds: M. J. Plavec, D. M. Popper and R. K. Ulrich (Reidel), 251

\bibitem[1999]{rich99}
Richichi, A., Fabbroni, L., Ragland, S., Scholz, M. 1999, A\&A, 344, 511

\bibitem[1992]{schaller92}
Schaller, G., Schaerer, D., Meynet G., Maeder, A. 1992, A\&AS, 96, 269

\bibitem[1997]{schmid97}
Schmid, H.M., 1997, in Physical processes in symbiotic binaries, ed. J. Miko{\l}ajewska 
(Copernicus  Found. for Polish Astronomy, Warsaw), 21

\bibitem[1980]{schneider80}
Schneider, D. P., Young, P. 1980, ApJ, 238, 946

\bibitem[2001]{schild01}
Schild, H., Dumm, T., M\"{u}rset, U., et al. 2001,
A\&A, 366, 972  (S01)

\bibitem[1983]{shafter83}
Shafter, A. W. 1983, ApJ, 267, 222

\bibitem[2000a]{skopal00}
Skopal, A., Djura\~sevi\'c, G., Jones, A., Rovithis-Livaniou, E., Rovithis, P. 2000a, MNRAS, 311, 225

\bibitem[2000b]{skopal00b}
Skopal, A., Pribulla, T., Wolf, M., Shugarov, S.Y., Jones, A. 2000b, Contrib. Astron. Obs. Skalnate Pleso, 30, 29

\bibitem[1983]{stone83}
Stone, R. P. S., Baldwin, J. A. 1983, MNRAS, 204, 347

\bibitem[1974]{th74}
Thackeray, A.D., Hutchings, J.B. 1974, MNRAS, 167, 319 (TH74)

\bibitem[1993]{winckel93}
Van Winckel, H., Duerbeck, H. W., Schwarz, H. E. 1993, A\&AS, 102, 401

\bibitem[1993]{vassi93}
Vassiliadis, E. \& Wood P. R. 1993, ApJ, 413, 641

\bibitem[2000]{wf01} 
Walder, R., Follini, D. 2000, in Thermal and Ionization Aspects of 
Flows from Hot Stars: Observations and Theory, ed. Henny J.G.L.M. Lamers, \& A. Sapar, ASP 
Conf. Ser., 204, 331

\end{thebibliography}
\end{document}